\documentclass[
prl,
twocolumn,
showpacs,
preprintnumbers,
amsmath]{revtex4}

\usepackage{graphicx}	% package for inserting EPS figures

\usepackage{amssymb}	% better font - can be commented out
\usepackage{amsmath}	% better font - can be commented out
\usepackage{mathptm}	% better font - can be commented out
\usepackage{color}	% to enable highligh changes in color

\newcommand{\be}{\begin{equation}}
\newcommand{\ee}{\end{equation}}
\newcommand{\bea}{\begin{eqnarray}}
\newcommand{\eea}{\end{eqnarray}}

\begin{document}       

% ----------------------- Title Page --------------------------------

\title{Does the 1/f frequency-scaling of brain signals \\
reflect self-organized critical states~?}

\author{C.~B\'edard$^{a,b}$, H.~Kr\"{o}ger$^{b}$,
A.~Destexhe$^{a,*}$} 

\affiliation{$^{a}$ {\small\sl Integrative and Computational
Neuroscience Unit (UNIC), CNRS, Gif-sur-Yvette, France} \\
$^{b}$ {\small\sl Department of Physics, Universit\'{e} Laval,
Qu\'{e}bec, Canada} \\ $*$ Corresponding Author, destexhe@iaf.cnrs-gif.fr}

% Alain Destexhe, \\ 
% UNIC, CNRS \\
% 1 Avenue de la Terrasse (Bat 33), \\ 
% 91198 Gif-sur-Yvette, France \\ 
% Tel: 33-1-69 82 34 35, Fax: 33-1-69 82 32 27 \\ 
% email: destexhe@iaf.cnrs-gif.fr

\date{\today}

% ----------------------- Abstract -----------------------------------

\begin{abstract}

Many complex systems display self-organized critical states
characterized by 1/f frequency scaling of power spectra.  Global
variables such as the electroencephalogram, scale as 1/f, which could
be the sign of self-organized critical states in neuronal activity. 
By analyzing simultaneous recordings of global and neuronal
activities, we confirm the $1/f$ scaling of global variables for
selected frequency bands, but show that neuronal activity is not
consistent with critical states.  We propose a model of 1/f scaling
which does not rely on critical states, and which is testable
experimentally.

\end{abstract}


\maketitle

% ----------------------- Text --------------------------------------

Self-organized critical states are found for many complex systems in
nature, from earthquakes to avalanches~\cite{Jensen98,Bak96}.  Such
systems are characterized by scale invariance, which is usually
identified as a power-law distribution of variables such as event
duration or the waiting time between events.  $1/f$ noise is usually
considered as a footprint of such systems~\cite{Jensen98}.  $1/f$
frequency scaling is interesting, because it betrays long-lasting
correlations in the system, similar to the behavior near critical
points.

Several lines of evidence point to the existence of such critical
states in brain activity.  Global variables, such as the
electroencephalogram (EEG) and magnetoencephalogram, display
frequency scaling close to $1/f$~\cite{Pritchard92,Novikov97}.  EEG
analysis~\cite{Link2001} and avalanche analysis of local field
potentials (LFPs) recorded {\it in vitro}~\cite{Beggs-Plenz} provided
clear evidence for self-organized critical states with power-law
distributions.  There is also evidence for critical states from the
power-law scaling of interspike interval (ISI) distributions computed
from retinal, visual thalamus and primary visual cortex
neurons~\cite{Teich97}.  In addition, model networks of neurons
indicate that critical states may be associated with frequency
scaling consistent with experiments~\cite{deArcangelis2006}. 
However, these are independent evidences from different preparations
and the link between $1/f$ frequency scaling of global variables and
the existence of critical states in neural activity has not been
firmly established.  Moreover, $1/f$ spectra are not necessarily
associated with critical states~\cite{DeLosRios99}, so it is not
clear if the intact and functioning brain operates in a way similar
to critical states.

\begin{figure}[h]
\begin{center}
\includegraphics[scale=0.8,angle=0]{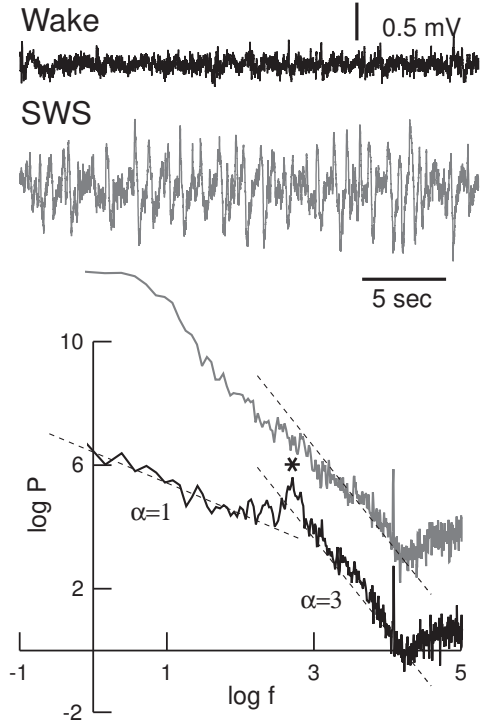}
\end{center}

\caption{Frequency scaling of local field potentials from cat
parietal cortex.  Top traces: LFPs recorded in cat parietal
{cortex} during wake and slow-wave sleep (SWS) states.  Bottom:
Power spectral density of LFPs, calculated from 55~sec sampled at
300~Hz {(150~Hz 4th-order low-pass filter)}, and represented in
log-log scale (dashed lines represent $1/f^\alpha$ scaling).  During
waking (black), the frequency band below 20~Hz scales approximately
as $1/f$ (*: peak at 20~Hz beta frequency), whereas the frequency
band between 20 and 65~Hz scales approximately as $1/f^3$.  During
slow-wave sleep (gray; displaced upwards), the power in the slow
frequency band is increased, and the $1/f$ scaling is no longer
visible, but the $1/f^3$ scaling at high frequencies remains
unaffected. PSDs were calculated over successive epochs of 32~sec,
which were averaged over a total period of 200~sec for Wake and
500~sec for SWS.}

\label{lfps}
\end{figure}

To attempt answering these questions, we first investigated if $1/f$
frequency scaling is present in global variables recorded close to
the underlying neuronal current sources {\it in vivo}.  We analyzed
cortical LFPs which were recorded within cerebral cortex using
bipolar extracellular high-impedance
microelectrodes~\cite{Destexhe99}.  Bipolar LFP recordings sample
relatively localized populations of neurons, as these signals can be
very different for electrodes separated by 1~mm
apart~\cite{Destexhe99}.  This stands in contrast with the EEG, which
samples much larger populations of neurons~\cite{Niedermeyer98} and
is recorded from the surface of the scalp using millimeter-scale
electrodes.  LFPs are subject to much less filtering compared to EEG,
because EEG signals must diffuse through various media, such as
cerebrospinal fluid, dura matter, cranium, muscle and skin.  Thus,
finding $1/f$ frequency scaling of bipolar LFPs would be a much
stronger evidence that this scaling reflects neuronal activities, as
these signals are directly recorded from within the neuronal tissue. 
Moreover, in order to distinguish state-dependent scaling properties,
we have compared recordings during wakefulness and slow-wave sleep in
the same experiments.  

Bipolar LFPs from cat parietal association cortex show the classic
landmarks of EEG signals in these states~\cite{Steriade03}, namely
during waking, LFPs are of low amplitude and very irregular
(Fig.~\ref{lfps}, top trace), and are dominated by beta frequencies
(around 20~Hz).  This pattern is also called ``desynchronized''
activity, and is typically seen during aroused states in the human
EEG~\cite{Niedermeyer98}.  During slow-wave sleep, LFPs display
high-amplitude slow-wave activity (Fig.~\ref{lfps}, middle trace),
similar to the ``delta waves'' of human sleep
EEG~\cite{Niedermeyer98}.  The power spectral density (PSD)
calculated from these LFPs typically shows a broad-band structure. 
During wakefulness, the PSD shows two different scaling regions,
according to the frequency band.  For low frequencies (between 1 and
20~Hz), the PSD scales approximately as $1/f$, whereas for higher
frequencies (between 20 and 65~Hz), the scaling is approximately of
$1/f^3$ (Fig.~\ref{lfps}, black PSD).  During slow-wave sleep, the
additional power at slow frequencies masks the $1/f$ scaling, but the
same $1/f^3$ scaling is present in the high-frequency band
(Fig.~\ref{lfps}, gray PSD).  The same behavior was observed for
other electrodes in the same experiment, and in three other animals
(not shown).  Thus, these results confirm that the $1/f$ frequency
scaling reported in the EEG~\cite{Pritchard92} is also present in
bipolar LFPs from cat association cortex, but only during waking and
for specific frequency bands.  

\begin{figure}[h]
\begin{center}
\includegraphics[scale=0.8,angle=0]{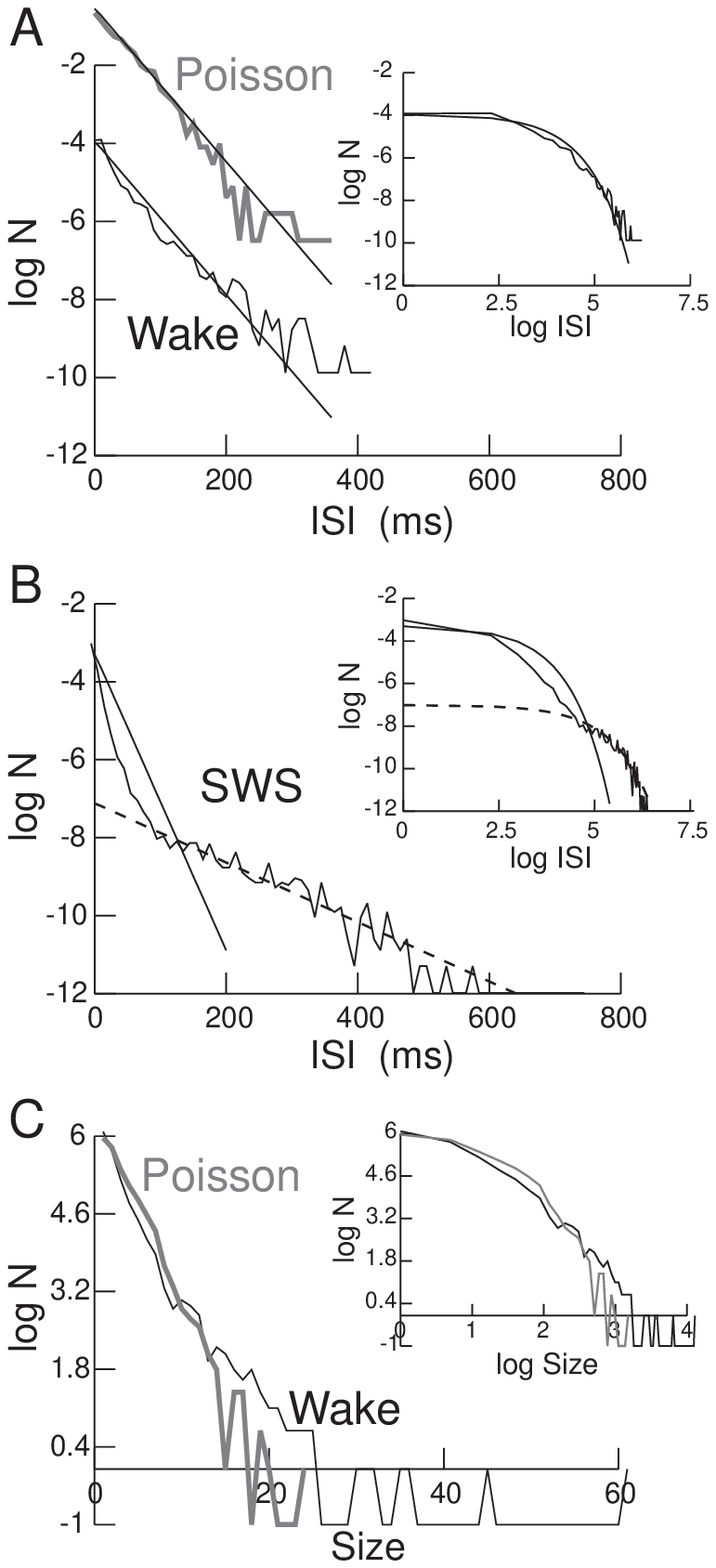}
\end{center}

\caption{Absence of power law distributions in neuronal activity. 
The logarithm of the distribution of interspike intervals (ISI)
during waking (Wake, A, 1951 spikes) and slow-wave sleep (SWS, B,
15997 spikes) is plotted as a function of ISI length, or log ISI
length (insets).  A poisson process of the same rate and statistics
is displayed in A (Poisson; gray curve displaced upwards for
clarity).  The exponential ISI distribution predicted by Poisson
processes of equivalent rates is shown as straight lines (smooth
curve in inset).  The dotted line in B indicates a Poisson process
with lower rate which fits the tail of the ISI distribution in SWS. 
C. Avalanche analysis realized by taking into account the
statistics from all simultaneously-recorded cells in Wake.  The
distribution of avalanche sizes scales exponentially (black curves),
similar to the same analysis performed on a Poisson process with same
statistics (gray curves).} 

\label{scal}
\end{figure}

To investigate whether this $1/f$ scaling is associated with
self-organized critical states, we first analyzed the ISI
distributions from neurons recorded in cat parietal cortex.  Unit
activity was recorded simultaneously with LFPs at 8 locations
separated by 1~mm~\cite{Destexhe99}.  The distribution of ISIs was
computed for individual neurons, and were represented in log-linear
scale (Fig.~\ref{scal}; log-log scale in insets).  For both
wakefulness and slow-wave sleep (Fig.~\ref{scal}A and B), the
distributions showed no evidence for power law behavior.  During
waking, the ISI distributions were close to exponentially-distributed
ISIs, as generated by Poisson stochastic process with same statistics
as the neurons analyzed (Fig.~\ref{scal}, Poisson).  For 22 neurons
recorded during the wake state, the Pearson coefficient was of
0.91~$\pm$~0.13 for exponential distribution fits, and of
0.86~$\pm$~0.16 for power-law distribution fits.  Taking only the
subset of 7 neurons with more than 2000 spikes, the fit was nearly
perfect for exponential distributions (Pearson coefficient of
0.999~$\pm$0.001).  However, during slow-wave sleep, there was a
marked difference between the experimental ISI and the corresponding
Poisson process (Fig.~\ref{scal}B).  In this state, neurons tended to
produce long periods of silences, which are related to EEG slow
waves~\cite{Destexhe99,Steriade03}, and which is visible as a
prominent tail of the distribution for large ISIs.  This tail was
well fit by a Poisson process of low rate (Fig.~\ref{scal}B, dashed
line).

To further check for criticality, we have performed an avalanche
analysis by taking into account the collective information from the
multisite  recordings.  We used the same method as for
ref.~\cite{Beggs-Plenz}, which amounts to detect clusters of
contiguous events separated by silences, by binning the system in
time windows of 1~ms to 16~ms~\cite{Beggs-Plenz}.  As there was no
evidence for any recognizable event in LFPs which could be taken as
avalanche (see Fig.~\ref{lfps}), we used the spike times among the
ensemble of simultaneously recorded neurons.  The distribution of
avalanche size does not follow power-law scaling (Fig.~\ref{scal}C,
black), but is closer to an exponential distribution as predicted by
Poisson processes (Fig.~\ref{scal}C, gray).  This analysis therefore
confirms the absence of avalanche dynamics in this
system{~\cite{Paczuski96}.}

\begin{figure}[h]
\begin{center}
\includegraphics[scale=0.4,angle=0]{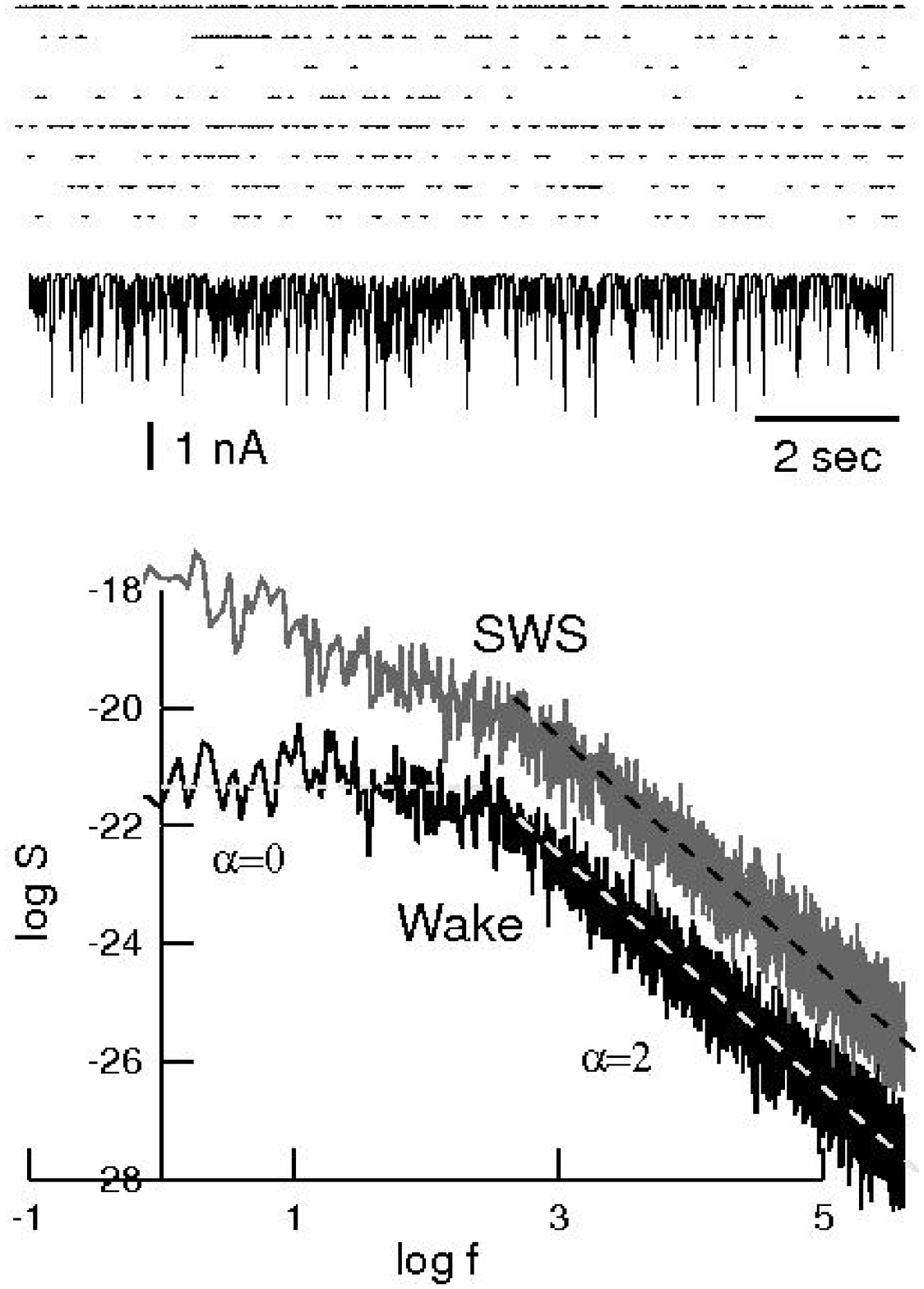}
\end{center}

\caption{Frequency scaling of synaptic currents reconstructed from
spike times.  Top traces: raster plot of spiking times of 8
multi-unit recordings in cat cortex during wakefulness (same
experiment as in Fig.~\ref{lfps}; data from ref.~\cite{Destexhe99}). 
Middle trace: total synaptic current obtained by convolving the spike
times with exponential relaxation processes ($\tau_s$=10~ms). 
Bottom: PSD of synaptic currents for wake (black) and slow-wave sleep
(SWS; PSD in gray displaced upwards for clarity);  dashed lines
represent $1/f^\alpha$ scaling.}

\label{curr}
\end{figure}

To explain the $1/f$ scaling of LFPs, we attempted to reconstruct
LFPs from unit activity.  Unit activity is displayed in
Fig.~\ref{curr} (top) for the same experiment as that of
Fig.~\ref{lfps}.  Because LFPs are generated primarily by synaptic
currents in neurons~\cite{Niedermeyer98,Nunez81}, and because
synaptic currents are very well modeled by simple exponential
relaxation processes~\cite{Destexhe98}, we modeled the synaptic
current from the following convolution~\cite{note-model}:

\be
  C(t) \ = \ \int_{-\infty}^\infty \ D(t') \ \exp[-(t-t')/\tau_s)] \ dt' ~ , 
\ee 
where $C(t)$ is the synaptic current and $D(t)$ is the ``drive''
signal which consisted in the experimentally-recorded spike trains.  
The PSD of the synaptic current is then given by: 
\be
  S(\omega) \ = \ |C(\omega)|^2 \ = \ 
  { |D(\omega)|^2 \over 1 + \omega^2 \tau_s^2} ~ .  \label{psd1}
\ee

The PSD of synaptic currents reconstructed from
experimentally-recorded spikes showed an approximate Lorentzian
behavior ($1/f^2$ scaling) during wakefulness (Fig.~\ref{curr},
Wake), as expected from the exponential nature of synaptic events. 
During slow-wave sleep, there was more power for slow frequencies,
but the $1/f^2$ scaling at high frequencies was still present
(Fig.~\ref{curr}, SWS).  The Lorentzian form of the PSD in
Fig.~\ref{curr} (Wake) shows that in the waking state,
$|D(\omega)|^2$ is approximately constant, therefore the drive $D(t)$
is statistically equivalent to a white noise process, consistent with
the apparent Poisson statistics of spikes identified in
Fig.~\ref{scal} (see also refs.~\cite{Softky93} for similar findings
in awake monkeys).  During slow-wave sleep, however, the deviation
from the Lorentzian suggests that $D(t)$ is a stochastic process
statistically different from white noise, and contains in addition
increased power at low frequencies, also consistent with the analysis
of Fig.~\ref{scal}.  

This model, however, does not yield PSD consistent with the $1/f$ and
$1/f^3$ scaling of LFPs shown in Fig.~\ref{lfps}.  Interestingly, the
scaling of this model is in $1/f^0$ or $1/f^2$ for the same frequency
bands that displayed $1/f$ or $1/f^3$ in LFPs, respectively.   Using
a similar convolution equation to model the PSD of LFPs
\be
  LFP(t) \ = \ \int_{-\infty}^\infty \ C(t') \ F(t-t') \ dt' ~ , 
\ee 
where $C(t)$ is the synaptic current source and $F(t)$ is a function
representing a filter.  As above, the PSD is given by
\be
  P(\omega) \ = \ |LFP(\omega)|^2 \ = \ |C(\omega)|^2 \ |F(\omega)|^2 
  ~ .  \label{psd2}
\ee

In this model, the frequency scaling of the PSD of both wakefulness
and slow-wave sleep LFPs in Fig.~\ref{lfps} can be explained by
assuming that the filter scales as $1/f$, or equivalently that
$|F(\omega)|^2 \ \sim \ 1 / \omega$.  In other words, this model can
explain qualitatively the $1/f$ and $1/f^3$ scaling of LFPs under the
condition that neuronal current sources are subject to an $1/f$
filter.  Such a filter is most likely due to the filtering of
extracellular currents through the tissue, before it reaches the
electrode~\cite{Bedard04}.

Finally, we provide an intuitive justification for this predicted
$1/f$ filter, as well as possible ways to test it experimentally. 
The $1/f$ filtering of extracellular media can be justified
intuitively by considering the complex structure of such media, and
in particular its spatial irregularity.  Extracellular space consists
of a complex arrangement of cellular processes of various size and
irregular shape, while the extracellular fluid represents only a few
percent of the available space~\cite{Peters91}.  The effect of a
current source in such media will be a combination of resistive
effects, due to the flow of current in the conductive fluids, and
capacitive effects, due to the high density of membranes (for a
theoretical treatment see refs.~\cite{Bedard04}).  Such a complex
arrangement of resistors and capacitors with random values is known
to produce an $1/f$ filter, as found for inhomogeneous
materials~\cite{Almond01}.  Although such materials are different
from the structure of biological media, it is plausible that similar
considerations may explain the $1/f$ filtering predicted here. 
Linear arrangements of RC circuits with random values (RC line) also
generates $1/f$ noise~\cite{Verveen74}.  Superposition of a large
number of exponential relaxation processes with different relaxation
rates can also generate $1/f$
scaling~\cite{Bernamont37,note-bernamont}.  {Understanding of
the $1/f$ filtering by extracellular media based on plausible
biophysical models is presently under investigation.  The predicted
$1/f$ filter} could also be tested experimentally by injecting white
noise currents (of amplitude comparable to neuronal current sources)
in extracellular space, and measuring the resulting field potential
at some distance from the injection site.  This measured LFP should
scale as $1/f$.

\

In conclusion, we have shown that the PSD of bipolar LFPs from cat
parietal cortex displays several scaling regions, as $1/f$ or $1/f^3$
depending on the frequency band and behavioral state.  By analyzing
neuronal unit activity from the same experiments, we did not see
evidence that this $1/f$ {scaling} is associated with critical
states.  Neither ISI distributions nor avalanche size distributions
display power-law scaling, but are rather consistent with Poisson
processes.  We provided an alternative explanation for $1/f$
frequency scaling which does not rely on critical states, but rather
stems from the filtering properties of extracellular media.  We gave
an intuitive explanation for a possible physical origin of such $1/f$
filtering, as well as a way to test it experimentally.  These results
may appear to contradict with previous evidence for critical states
in vitro~\cite{Beggs-Plenz} or in the early visual system in
vivo~\cite{Teich97}.  However, the absence of critical states
reported here may instead reflect fundamental differences between
association cortex and other structures more directly related to
sensory inputs.  Future work should clarify why different structures
show different scaling, and what implications it may have for brain
dynamics and coding.

% --------------- Acknowledgments and references -------------------

\

The experimental data analyzed in this article were obtained with
Drs.\ Diego Contreras and Mircea Steriade, and were published
previously~\cite{Destexhe99}.  We are grateful for support by NSERC
Canada (H.K.), CNRS, the European Commission (FET program), and the
HFSP program (A.D.).


\begin{thebibliography}{99}
\bibitem{note-model} Nearly identical results were obtained using a
biophysical model based on a passive membrane equation where
experimentally-recorded spikes triggered conductance-based synaptic
inputs.
\bibitem{note-bernamont} This mechanism is however not plausible,
because it would require a uniform distribution of decay times of
synaptic currents across a large range of values, whereas only a few
receptor types are present in biological tissue, with well-defined
decay times.
\end{thebibliography}
\end{document}